\documentstyle[aps,twocolumn]{revtex}
\title{Quantum information and physics: some future
directions\thanks{CALT-68-2219}}

\author{John Preskill,\thanks{\tt preskill@theory.caltech.edu}\\
{\sl Lauritsen Laboratory of High Energy Physics\\
California Institute of Technology\\
Pasadena, CA 91125, USA}}
\date{April, 1999}

\begin{document}

\maketitle

\begin{abstract}
I consider some promising future directions for quantum information theory that
could influence the development of 21st century physics. Advances in the theory
of the distinguishability of superoperators may lead to new strategies for
improving the precision of quantum-limited measurements. A better grasp of the
properties of multi-partite quantum entanglement may lead to deeper
understanding of strongly-coupled dynamics in quantum many-body systems,
quantum field theory, and quantum gravity.

\end{abstract}

\section{Introduction}

With the discovery of an apparent separation between the classical and quantum
classifications of computational complexity \cite{shor}, and of fault-tolerant
schemes for quantum computation \cite{shor_ft}, quantum information theory has
earned a lasting and prominent place at the foundations of computer science.
But at present this discipline seems rather isolated from most of the rest of
physics.  Will this change in the future?  How might it change?

One view is that thinking about information theory will lead us to a deeper
understanding of the foundations of quantum mechanics.  This vision has been
vividly expressed by John Wheeler \cite{wheeler}; Bill Wootters \cite{wootters}
and Chris Fuchs \cite{fuchs} have been among its particularly eloquent
spokespersons.  But I am not convinced in my heart that we are supposed to
understand the foundations of quantum mechanics much better than we currently
do.  So I prefer to look in a different direction to anticipate where quantum
information may have an impact on physics.

What I tend to find most exciting in science are ideas that can build bridges
across the traditional boundaries between disciplines. Perhaps that is why I
find quantum computation appealing --- it has established an unprecedentedly
deep link between the foundations of computer science and the foundations of
physics.  Truly great ideas in science tend to have broad consequences that
can't be anticipated easily.

Now the quantum information community is sitting atop two ideas with potential
for greatness: quantum computation and quantum error correction.  I'd like to
suggest two directions in which quantum information theory might evolve in the
future that could lead to broad and exciting consequences for other subfields
of physics. These are:
\begin{description}
\item{1.} {\bf Precision measurement.} \newline
Our deepening understanding of quantum information may lead to new strategies
for pushing back the boundaries of quantum-limited measurements.  Quantum
entanglement, quantum error correction, and quantum information processing
might all be exploited to improve the information-gathering capability of
physics experiments.
\item{2.} {\bf Many-body quantum entanglement.} \newline
The most challenging and interesting problems in quantum dynamics involve
understanding the behavior of strongly-coupled many-body systems --- systems
with many degrees of freedom that undergo large quantum fluctuations.  Better
ways of characterizing and classifying the features of many-particle
entanglement may lead to new and more effective methods for understanding the
dynamical behavior of complex quantum systems.
\end{description}

\section{Quantum information theory and precision measurement}

The connections between quantum information and precision measurement are
explored in a separate article \cite{childs}, which I will only summarize here.

My own interest in the quantum limitations on precision measurement has been
spurred in part by Caltech's heavy involvement in the LIGO project, the Laser
Interferometer Gravitational-Wave Observatory \cite{ligo}. LIGO is scheduled to
begin collecting data in 2002, and a major upgrade is planned for two years
later, which will boost the optical power in the interferometer and improve the
sensitivity.  In its most sensitive frequency band, the LIGO II observatory
will actually be operating at the standard quantum limit (SQL) for detection of
a weak classical force by monitoring a free mass. (In this case, the SQL
corresponds to a force that nudges an 11 kg mass by about $10^{-17}$ cm at a
frequency of 100 Hz.)

Then within another 4 years (by 2008), another upgrade is expected, which will
boost the sensitivity in the most critical frequency band beyond the SQL.  Even
an improvement by a factor of two can have a very significant payoff, for a
factor of two in sensitivity means a factor of 8 in event rate. But the design
of the LIGO III detection system is still largely undecided --- clever
innovations will be needed.  So Big Science will meet quantum measurement in
the first decade of the new century, and ideas from quantum information theory
may steer the subsequent developments in detection of gravitational waves and
other weak forces.

I learned the right way to think about the quantum limits on measurement
sensitivity from Hideo Mabuchi \cite{mabuchi} --- in a quantum measurement, a
classical signal is conveyed over a quantum channel.\footnote{Of course,
connections between quantum information theory and precision measurement have
been recognized by many authors. Especially relevant is the work by Wootters
\cite{wootters}, by Braunstein\cite{braun}, and by Braunstein and Caves
\cite{braunstein} on state distinguishability and parameter estimation, and by
Braginsky and others \cite{braginsky} on quantum nondemolition measurement.}
Nature sends us a message, a weak classical force, that can be regarded as a
classical parameter appearing in the Hamiltonian of the apparatus (or more
properly, if there is noise, a master equation).  The apparatus undergoes a
quantum operation $\$(a)$, and we are to extract as much information as we can
about the parameter(s) $a$ by choosing an initial preparation of the apparatus,
and a POVM to read it out.  Quantum information theory should be able to
provide a theory of the {\sl distinguishability of superoperators}, a measure
of how much information we can extract that distinguishes one superoperator
from another, given some specified resources that are available for the
purpose.  This distinguishability measure would characterize the inviolable
limits on measurement precision that can be achieved with fixed resources.

I don't know exactly what shape this nascent theory of the distinguishability
of  superoperators should take, but there are already some highly suggestive
hints that progress in quantum information processing can promote the
development of new strategies for performing high-precision measurements.

\subsection{Superdense coding: improved distinguishability through
entanglement}

A watchword of quantum information theory is: ``Entanglement is a Useful
Resource.'' It should not be a surprise if entanglement can extend the
capabilities of the laboratory physicist.

For example, the phenomenon of superdense coding illustrates that shared
entanglement can enhance classical communication between two parties
\cite{wiesner}.  The same strategy can sometimes be used to exploit
entanglement to improve the distinguishability among Hamiltonians (an idea
suggested by Chris Fuchs \cite{fuchs_sd}).  Suppose I wish to observe the
precession of spin-$1/2$ objects to determine the value of an unknown magnetic
field.  If two spins are available, one way to estimate the value of the
unknown field is to allow both spins to precess in the field independently, and
then measure them separately. An alternative method is to prepare an entangled
Bell pair, expose one of the two spins to the magnetic field while the other is
carefully shielded from the field, and finally carry out a collective Bell
measurement on the pair.   It turns out that in many cases (for example when we
have no {\sl a priori} knowledge about the field direction), the entangled
strategy extracts more information about the unknown field than the strategy in
which uncorrelated spins are measured one at a time \cite{childs}. This
separation still holds even if we allow the unentangled strategy to be {\sl
adaptive}; that is, even if the outcome of the measurement of the first spin is
permitted to influence the choice of the measurement that is performed on the
second spin.

\subsection{Grover's database search: improved distinguishability through
driving}

An important paradigm emerging from the recent studies of quantum algorithms is
Grover's method for rapidly searching an unsorted database \cite{grover}. Farhi
and Gutmann \cite{farhi} observed that Grover's algorithm may be interpreted as
a method for improving the distinguishability of a set of Hamiltonians by
adding a controlled driving term.

In the formulation they suggested, the Hamiltonian acting in an $N$-dimensional
Hilbert space is known to be one of the operators
\begin{equation}
H_x= E|x\rangle\langle x|~,
\end{equation}
where $\{|x\rangle, x=0,1,\dots,N-1\}$ is an orthonormal basis.  We may gain
information about the value of $x$ by preparing states, allowing them to evolve
under $H_x$ for a while, and then measuring suitable observables. But
determining the value of $x$ by this strategy requires a total time of order
$N$. A more effective strategy is to modify the Hamiltonian by adding a
controlled driving term
\begin{equation}
H_D= E|s\rangle\langle s| ~,
\end{equation}
where $|s\rangle=N^{-1/2}\sum_{y=0}^{N-1}|y\rangle$, so that the full
Hamiltonian becomes $H'_x=H_x+H_D$.
If the initial state $|s\rangle$ is prepared and allowed to evolve under $H'_x$
for a time $T=\pi\sqrt{N}/2E$, then an orthogonal measurement in the
$\{|x\rangle\}$ basis will reveal the true identity of the Hamiltonian.  The
time required is of order $\sqrt{N}$; this is Grover's quadratic speed-up.

In this Grover-Farhi-Gutmann problem, there is a sense in which an optimal
measurement procedure is known: Just as the Grover iteration allows one to
identify a marked state with a minimum number of queries to the oracle
\cite{bbbv}, the Grover perturbation allows us to identify the actual
Hamiltonian in the minimal elapsed time (asymptotically for large $N$).

Grover's algorithm presumes the existence of a quantum oracle that can reply to
coherent queries. In an algorithmic setting, the oracle may be regarded as a
quantum circuit that can be executed repeatedly.  In experimental physics, the
quantum oracle is Nature, whose secrets we are eager to expose.  The
experimenter is challenged to find the most effective (and practical!) way to
query Nature and learn her Truths.

\subsection{Semiclassical quantum Fourier transform as adaptive phase
measurement}

Shor's quantum factoring algorithm \cite{shor}, which apparently achieves an
exponential speed-up relative to classical algorithms, is based on the
efficient quantum Fourier transform (QFT). Fourier analysis is a versatile tool
in the laboratory, so we might expect the fast QFT to have important
applications to physics.

One example could be the high-precision measurement of a frequency, like the
energy splitting between the ground state and an excited state of an atom
\cite{childs}.  As Cleve {\it et al.} \cite{cleve} have emphasized, the QFT can
be viewed as a procedure for estimating an unknown phase. With a quantum
computer, we could execute the quantum Fourier transform on $n$ two-level
atoms, and then read out a result by measuring the internal state of each atom.
 If losses are negligible, the measurement outcomes provide an estimate of the
frequency to an accuracy of order $2^{-n}$.  This procedure makes optimal use
of an essential resource (the number of atoms measured), in that about one bit
of information about the frequency is acquired in each binary measurement.

In fact, the complexity of the quantum information processing needed to execute
this protocol is modest. In its ``semiclassical'' implementation proposed by
Griffiths and Niu \cite{griffiths}, the QFT is an {\sl adaptive} procedure for
phase estimation that makes use of the information collected in previous
measurements to extract the best possible information from subsequent
measurements.  Less significant bits of the phase are measured first, and the
measurement results are used to determine what single-qubit phase rotations
should be applied to other qubits to extract the more significant bits more
reliably.  In conventional Ramsey spectroscopy, these single-qubit
transformations are applied simply by prescribing the proper time interval
between the Ramsey pulses.

\bigskip

These and other related examples give strong hints that ideas emerging from the
theory of quantum information and computation are destined to profoundly
influence the experimental physics techniques of the future.

\section{\bf Many-body entanglement and strongly-coupled quantum physics}

\subsection{Some signposts in Hilbert space}

The most challenging and interesting problems in quantum mechanics concern
many-body systems with strong quantum fluctuations.  An important goal is to
understand the dynamics of such systems, but it is not easy.  Indeed, it is
largely because strongly-coupled quantum dynamics is so difficult to understand
that we want so badly to build a quantum computer \cite{feynman}!

I expect that, short of building a full-blown quantum simulator, there are many
possible theoretical advances that potentially could enhance our understanding
of strongly-coupled systems, including advances that could emerge from the
theory of quantum information.  A central task of quantum information theory
has been to characterize and quantify the entanglement of multipartite systems.
 Up until now, most attention has focused on systems divided into a small
number of parts (like two\footnote{See \cite{ibm}, for example.}), but also of
great importance are the properties of $n$-body entanglement in the limit of
large $n$. Studies of these properties may give us some guidance concerning
what quantum simulation problems are genuinely computationally difficult, and
may suggest to experimenters what kinds of systems are most likely to exhibit
qualitatively new phenomena.

Hilbert space is a big place \cite{caves_fuchs}, and so far we have become
familiar with only a tiny part of it.  In its unexplored vastness, there is
sure to be exciting new physics to discover.  But much of Hilbert space is
bound to be very boring indeed, so we will need some clear signposts to show
the way to the exotic new phenomena.

It is truism (but still profoundly true!) that More is Different
\cite{anderson}.  So many of the collective phenomena exhibited by many-body
systems (crystals, phase transitions, superconductivity, fractional quantum
Hall effect, $\dots$) would be exceedingly hard to predict from first
principles.  That's good news for experimenters --- marvelous things could
happen in many-body systems that we have been unable to anticipate. But it is
easier to find something new when theory can provide some guidance.

\subsection{Quantum error-correcting codes}

A prototype for many-body entanglement has been developed in the past few
years: the quantum error-correcting codes \cite{qec}.  For example, a
(nondegenerate) code that can correct any $t$ errors in a block of $n$ qubits
has the property that no information resides in any set of $2t$ qubits chosen
from the block -- the density matrix of the $2t$ qubits is completely random.
Information {\sl can} be encoded in the block, but the encoded information has
a {\sl global} character; there is no way to access any information at all by
looking at only a few qubits at a time.

For example, associated with the familiar five-qubit code \cite{ibm,five_qubit}
that can protect a single encoded qubit from an error afflicting any of the
five qubits in the code block, there is a maximally entangled six-qubit pure
state.  This state has the property that if we trace over any three of the
qubits, the density matrix of the remaining three is a multiple of the
identity.  It has been recognized only rather recently how unusual this state
is  \cite{gf4}:  there exist no $2n$-qubit states with $n$ larger than three
such that tracing over half of the qubits leaves the other half in a completely
random state.\footnote{But there {\sl are} such maximally entangled states with
more than six parts if each part is a higher-dimensional system rather than a
qubit \cite{gottesman}.}

Asymptotically, we don't know precisely ``how entangled''  an $n$-qubit state
can be, but there are useful upper and lower bounds.  For large $n$, the number
$s$ of qubits such that the density matrix for any $s$ of the $n$ is random,
{\sl must} satisfy $s/n < 1/3$ \cite{rains}.  On the other hand, states with
this property are known to exist for $s/n \le .1893\dots$ \cite{att}. Somewhere
between $1/3$ and $.1893$, there is a critical value that has not yet been
pinned down. These upper and lower bounds are instructive examples of
interesting results regarding multi-body entanglement that have emerged from
the study of quantum error-correcting codes.

\subsection{Classes of entangled states}

This kind of global encoding of information is actually found in some systems
that can be realized in the laboratory, such as systems that exhibit the
fractional quantum Hall effect \cite{stone}, or certain kinds of frustrated
antiferromagnets.  These systems have in common that the microscopic degrees of
freedom are locally ``frustrated'' -- that is, they are unable to find a
configuration that satisfactorily minimizes the local energy density.  In
response, the system seeks an unusual collective state that relieves the
frustration, a state such that the microscopic degrees of freedom are
profoundly entangled.  Condensed matter physicists have found useful ways to
characterize the global properties of the entanglement that results.

For example, in the case of a two-dimensional system, we may consider how the
ground state degeneracy of the system behaves on a topologically nontrivial
surface in the thermodynamic limit.  As Wen \cite{wen} emphasized, in
fractional quantum Hall systems the degeneracy increases with the genus (number
of handles) of the surface as
\begin{equation}
{\rm ground ~state ~degeneracy}\sim (A)^{\rm genus} ~.
\end{equation}
This dependence arises from the ``winding'' of entanglement around the handles
of the surface, and the value of $A$ distinguishes qualitatively different
types of entangled states that must be separated from one  another by phase
boundaries. Just such a topological degeneracy is exploited in the ingenious
quantum error-correcting codes constructed by Alexei Kitaev \cite{kitaev}. A
closely related observation is that in a two-dimensional system with a
boundary, there can be excitations confined to the boundary, and the properties
of these edge excitations reflect the nature of the entanglement in the bulk
system \cite{wen_edge}.

I am hopeful that quantum information theory may lead to other as yet unknown
ways to characterize the entangled many-body ground states of condensed matter
systems, which may suggest new types of collective phenomena.  We should also
advance our understanding of how the profoundly entangled systems that Nature
already provides might be exploited for stable storage of quantum information.

\subsection{Information and renormalization group flow}

The renormalization group (RG), one of the most profound ideas in science, is
another topic that might be profoundly elucidated by an information-theoretic
approach. Especially in the hands of Ken Wilson \cite{epsilon}, the RG spawned
one of the central unifying insights of modern physics, that of {\sl
universality} --- physics at long distances can be quite insensitive to the
details of physics at much shorter distances.  Indeed, for the purpose of
describing the long-distance physics, all of the short-distance physics can be
absorbed into the values of the parameters of an {\sl effective field theory},
where the number of parameters needed is modest if we are content with
predictions to some specified accuracy.

So it is that physics is possible at all. Fortunately, it is not necessary to
grasp all the subtleties of quantum gravity at the Planck scale to understand
(say) the spectrum of the hydrogen atom in great detail!

The renormalization group describes how a quantum field theory ``flows'' as we
``integrate out'' short distance physics, obtaining  a new theory with a
smaller value of the ultraviolet momentum cutoff $\Lambda$.  ``Universal''
features are associated with the ``fixed'' points in the space of theories
where the flow is stationary.  In the neighborhood of each fixed point are a
finite number of independent directions in theory space along which the flow is
repelled by the fixed point, the ``relevant'' directions of flow.

Each fixed point provides a potential description of physics in the far
infrared, with the number of free (``renormalized'') parameters in the
description given by the number of relevant directions of flow away from the
fixed point.  Infrared theories with more parameters are less generic, in the
sense that more ``bare'' parameters in the microscopic Hamiltonian of the
system need to be carefully tuned in order for the flow to avoid all relevant
directions and hence carry the theory to the vicinity of the fixed point.
Typically, RG flow will carry a theory from the vicinity of a less generic
fixed point toward the vicinity of a more generic fixed point.

Now there is at least a heuristic sense in which information is lost as a
theory flows along an RG trajectory --- the infrared theory ``forgets'' about
its ultraviolet origins.  One of the most intriguing challenges at the
interface of physics and information is to make this connection more
concrete.\footnote{For an interesting recent attempt, see \cite{brody}.}  Can
we quantify how much information is discarded when a theory flows from the
vicinity of one fixed point to the vicinity of another?

The proposal that an effective theory forgets more and more about its
microscopic origins under RG flow leads to a robust expectation. RG flow should
be a {\sl gradient} flow: it always runs downhill (toward ``less
information''), and never uphill (toward ``more information'').  Indeed, this
property {\sl does} hold for translationally invariant and relativistically
invariant quantum field theories in one spatial dimension.  Zamalodchikov's
$c$-{\sl theorem} \cite{ctheorem} identifies a function $C$ of the parameters
in the Hamiltonian that can extracted from the two-point correlation function
of the conserved energy-momentum tensor, and shows that $C$ is non-increasing
along an RG trajectory. At a fixed point, the quantity $C$ coincides with the
central charge $c$ that characterizes the representation of the conformal
algebra according to which the fields of the fixed-point theory transform. Last
year, an extension of this result to higher even-dimensional spacetimes was
reported \cite{c4d} (following a suggestion by Cardy \cite{cardy_c}).

It seems natural that the Zamolodchikov $C$-function should have a sharp
interpretation relating it to loss of information along the flow, but none such
is known (at least to me).
A more precise information-theoretic interpretation of RG flow might guide the
way to more general formulations of the $c$-theorem, applicable for example to
theories in odd-dimensional spacetimes and to theories with less symmetry.  And
it might enrich our understanding of the  classification of fixed-point
theories and the general structure of renormalization group flow.

\subsection{Bulk-boundary interactions}

If the information-theoretic foundations underlying the $c$-theorem continue to
prove elusive, there is another related problem that might turn out to be more
tractable. It is known that a one-dimensional system  with a {\sl boundary}
(like a semi-infinite antiferromagnetic spin chain) can sometimes exhibit an
anomalous zero-temperature entropy.  The entropy has a piece proportional to
the length of the chain that vanishes as $T\to 0$, but there is also a
length-independent contribution that is nonvanishing at zero temperature
(discovered by Cardy \cite{cardy} and by Affleck and Ludwig \cite{affleck}).
Ordinarily, we expect that zero-temperature entropy has an interpretation in
terms of ground-state degeneracy, but in these systems (which have no mass gap,
so that the ground-state degeneracy becomes a subtle concept in the
thermodynamic limit), $g=e^{S(T=0)}$ is not an integer; hence the
interpretation of the entropy is obscure.

A fascinating feature is that the ``ground-state degeneracy'' $g$ is  a {\sl
universal} property --- in the vicinity of an RG fixed point, its value is
insensitive to the ultraviolet details (the microscopic interactions among the
spins in the chain). Furthermore, there is evidence for a $g$-{\sl theorem};
$g$ has a smaller value at more generic fixed points and a larger value at less
generic fixed points \cite{affleck}.

The $g$-theorem, like the $c$-theorem, invites an interpretation in terms of
loss of information along an RG trajectory.  But I am hopeful that the
information-theoretic origin of the $g$-theorem may turn out to be easier to
understand.  Upon hearing of entropy at zero temperature, a quantum information
theorist's ears prick up -- it sounds like entanglement. It is tempting to
interpret the entropy as arising from entanglement of degrees of freedom
isolated at the boundary of the chain with degrees of freedom that reside in
the bulk.  So far, I have been unable to find a precise interpretation of this
sort, but I still suspect that it could be possible.

\subsection{Holographic universe}  While on the subject of bulk-boundary
interactions, I should mention the most grandiose such interaction of all.  A
new view of the quantum mechanics of spacetime is emerging from recent work in
string theory, according to which the quantum information encoded in a spatial
volume can be read completely on the surface that bounds the volume (``the
holographic principle'') \cite{thooft}.  This too has a whiff of entanglement
-- for we have seen that in a profoundly entangled state the amount of
information stored locally in the microscopic degrees of freedom can be far
less than we would naively expect.  (Think of a quantum error-correcting code,
in which the encoded information may occupy a small ``global'' subspace of a
much larger Hilbert space.) The holographic viewpoint is particularly powerful
in the case of the quantum behavior of a black hole.  The information that
disappears behind the event horizon can be completely encoded on the horizon,
and so can be transferred to the outgoing Hawking radiation \cite{hawking} that
is emitted as the black hole evaporates.  This way, the evaporation process
need not destroy any quantum information.

As the evidence supporting the holographic principle mounts \cite{maldacena},
an unsettling question becomes more deeply puzzling:  If quantum information
can be encoded completely on the boundary, why does physics seem to be local?
It's strange that I imagine that I can reach out and embrace you, when we are
both just shadows projected on the wall. Perhaps as the tools for analyzing
many-body entanglement grow more powerful, we can begin to grasp the origin of
the persistent illusion that physics is founded on the locality of
spacetime.\footnote{A different possible connection between quantum
error-correcting codes and the black-hole information puzzle was suggested in
\cite{preskill_ft}.}

\section{Conclusions}
In the future, I expect quantum information to solidify its central position at
the foundations of computer science, and also to erect bridges that connect
with precision measurement, condensed matter physics, quantum field theory,
quantum gravity, and other fields that we can only guess at today.  I have
identified two general areas in which I feel such connections may prove to be
particularly enlightening. Progress in quantum information processing may guide
the development of new ideas for improving the information-gathering
capabilities of physics experiments. And a richer classification of the phases
exhibited by highly entangled many-body systems may deepen our appreciation of
the wealth of phenomena that can be realized by strongly-coupled quantum
systems.

\acknowledgments
My work on the applications of quantum information theory to quantum-limited
measurements has been in collaboration with Andrew Childs and Joe Renes
\cite{childs}. I'm very grateful to Hideo Mabuchi for stimulating my interest
in that subject, and to Dave Beckman and Chris Fuchs for their helpful
suggestions. I have also benefitted from discussions about precision
measurement with Constantin Brif, Jon Dowling, Steven van Enk, Jeff Kimble,
Alesha Kitaev, and Kip Thorne.  I thank Michael Nielsen for emphasizing the
relevance of quantum information in quantum critical phenomena, Ian Affleck for
enlightening correspondence about conformal field theory, Anton Kapustin for a
discussion about Ref. \cite{c4d}, Dorje Brody for informing me about Ref.
\cite{brody}, and Curt Callan for encouragement. Finally, I am indebted to Ike
Chuang for challenging me to speculate about  the future of quantum information
theory. This work has been supported in part by the Department of Energy under
Grant No. DE-FG03-92-ER40701, and by DARPA through the Quantum Information and
Computation (QUIC) project administered by the Army Research Office under Grant
No. DAAH04-96-1-0386.

\end{document}